# Azimuthally unidirectional transport of energy in magnetoelectric fields: Topological Lenz's effect

R. Joffe[1,2], E. O. Kamenetskii[1], and R. Shavit[1]

[1] Microwave Magnetic Laboratory,
Department of Electrical and Computer Engineering,
Ben Gurion University of the Negev, Beer Sheva, Israel

[2] Department of Electrical and Electronics Engineering, Shamoon College of Engineering, Beer Sheva, Israel

January 17, 2017

**Abstract**

Magnetic-dipolar modes (MDMs) in a quasi-2D ferrite disk are microwave energy-eigenstate oscillations with topologically distinct structures of rotating fields and unidirectional power-flow circulations. At the first glance, this might seem to violate the law of conservation of an angular momentum, since the microwave structure with an embedded ferrite sample is mechanically fixed. However, an angular momentum is seen to be conserved if topological properties of electromagnetic fields in the entire microwave structure are taken into account. In this paper we show that due to the topological action of the azimuthally unidirectional transport of energy in a MDM-resonance ferrite sample there exists the opposite topological reaction on a metal screen placed near this sample. We call this effect topological Lenz's effect. The topological Lenz's law is applied to opposite topological charges: one in a ferrite sample and another on a metal screen. The MDM-originated near fields – the magnetoelectric (ME) fields – induce helical surface electric currents and effective charges on a metal. The fields formed by these currents and charges will oppose their cause.

## I.  INTRODUCTION

In dielectric microcavities, when sizes are comparable with a wavelength inside a dielectric material, the rotation modifies the cavity resonances. The electromagnetic fields in a rotating resonant microcavity are subject to the Maxwell equations generalized to a noninertial frame of reference in uniform rotation. Due to a non-electromagnetic torque, the path length of clockwise (CW) propagating light is different from that of counter-clockwise (CCW) propagating light. As a result, one has phase difference or frequency splitting between counter propagating beams in a loop. This is the Sagnac effect in optical microcavities [1, 2]. The phase and frequency difference between CW and counter-clockwise CCW propagating beams are proportional to the angular velocity. In the case of the two-dimensional (2D) disk dielectric resonator rotating at angular velocity $\Omega$, the resonances can be obtained by solving the following stationary wave equation [2]:

$$\frac{\partial^2 \varphi}{\partial r^2} + \frac{1}{r}\frac{\partial \varphi}{\partial r} + \frac{1}{r^2}\frac{\partial^2 \varphi}{\partial \theta^2} + 2ik\frac{\Omega}{c}\frac{\partial \varphi}{\partial \theta} + n^2 k^2 \varphi = 0, \qquad (1)$$

where $\varphi$ is an electric field component. The solutions for $\varphi(r,\theta)$ is given as $\varphi(r,\theta) = f(r)e^{im\theta}$, where $m$ is an integer. When the disk cavity is rotating, the wave function is the rotating wave $J_m(K_m r)e^{im\theta}$, where $K_m^2 = n^2 k^2 - 2k(\Omega/c)m$. The frequency difference between the counter-propagating waves is equal to $\Delta\omega = 2(m/n^2)\Omega$. For a given direction of rotation, the CW and CCW waves inside a dielectric cavity experience different refraction index $n$. Their azimuthal numbers are $\pm m$.

In the case of effects in rotating magnetic samples, very specific properties of geometrical phases become evident. In Ref. [3] it was shown that the sample rotation induces frequency splitting in nuclear-quadrupole-resonance (NQR) spectra. For the rotation frequency much less than the NQR characteristic frequency, the observed splitting in this magnetic resonance experiment can be interpreted in two ways: (*a*) as a manifestation of Berry's phase, associated with an adiabatically changing Hamiltonian, and (*b*) as a result of a fictitious magnetic field, associated with a rotating-frame transformation. The main standpoint is that the dynamical phase evolution in a magnetic structure is unaffected by the rotational motion and any rotational effects must arise from Berry's phase. Another studies on the rotational effects in magnetic samples are devoted to rotational Doppler effect observed in ferromagnetic-resonance (FMR) rotating magnetic samples. Frequencies of the FMR are typically in the gigahertz range, which is far above achievable angular velocities of mechanical rotation of macroscopic magnets. In [4, 5] it was discussed that a rotating ferromagnetic nanoparticle (having a very small moment of inertia) can be a resonant receiver of the electromagnetic waves at the frequency of the FMR. If such a receiver is rotating mechanically at an angular velocity $\Omega$, one has splitting of the FMR frequency. The frequency of the wave perceived by the receiver equals $\omega^{(n)} = \omega_{FMR} - \Omega$, where $\omega_{FMR}$ is the frequency of the ferromagnetic resonance, $\Omega = \frac{n\hbar}{I}$, $I$ is the moment of inertia of a nanoparticle, and $n = 0, \pm 1, \pm 2, \pm 3,...$ The experimental results for the rotational motion of solid ferromagnetic nanoparticles confined inside polymeric cavities give evidence for the quantization [5].

While in the above studies, specific wave effects are observed due to mechanical rotation of samples, there exists evidence for azimuthally unidirectional wave propagation in mechanically fixed magnetic samples. The multiresonance splitting in the FMR spectra was observed in mechanically non-rotating macroscopic quasi-2D ferrite disks with magnetic-dipolar-mode (MDM) (or magnetostatic-wave (MS-wave)) oscillations. Such oscillations in a ferrite disk are rotating eigenmodes with azimuthally unidirectional transport of energy [6 – 11]. In the MDM ferrite disks, the Berry's phase is generated from the broken dynamical symmetry [6]. It was shown that when a ferrite disk is loaded by a dielectric sample, the ferrite is under the reaction torque and the magnetization motion is characterized by a fictitious magnetic field [10, 11].

MDM oscillations in small (with sizes much less than the microwave radiation wavelength) ferrite spheres excited by external microwave fields were experimentally observed, for the first time, by White and Solt in 1956 [12]. Afterward, experiments with small ferrite disks revealed unique spectra of such oscillations. While in a case of a small ferrite sphere one observes only a few and very wide absorption peaks of MDM oscillations, for a small quasi-2D ferrite disk there is a multiresonance (atomic-like) spectrum with very sharp resonance peaks [13 – 15]. Analytically, it was shown [16, 17] that, contrary to spherical geometry of a ferrite particle analyzed in Ref. [18], the quasi-2D geometry of a ferrite disk gives the Hilbert-space energy-state selection rules for MDM spectra. MDM oscillations in a quasi-2D ferrite disk are macroscopically quantized states. Long range dipole–dipole correlation in position of electron spins in a ferromagnetic sample can be treated in terms of collective excitations of the system as a whole. When a ferrite-disk resonator with an energy spectrum is subjected to a weak action by an external microwave field, its energy levels do not change (change slightly, to be more precise). However, due to such a weak external action, the entire structure of precessing elementary magnetic dipoles acquires an orbital angular momentum. The MS-potential wave function in a ferrite disk with the disk axis oriented along $z$ axis, is written as [6 – 11]:

$$\psi = C\xi(z)\tilde{\varphi}(r,\theta), \qquad (2)$$

where $\tilde{\varphi}$ is a dimensionless membrane function, $r$ and $\theta$ are in-plane coordinates of a cylindrical coordinate system, $\xi(z)$ is a dimensionless function of the MS-potential distribution along $z$ axis, and $C$ is a dimensional amplitude coefficient. Being the energy-eigenstate oscillations, the MDMs in a ferrite disk also are characterized by topologically distinct structures of the fields. This becomes evident from the boundary condition on a lateral surface of a ferrite disk of radius $\mathcal{R}$, written for a membrane wave function [6 – 11]:

$$\mu\left(\frac{\partial \tilde{\varphi}}{\partial r}\right)_{r=\mathcal{R}^-} - \left(\frac{\partial \tilde{\varphi}}{\partial r}\right)_{r=\mathcal{R}^+} + i\frac{\mu_a}{\mathcal{R}}\left(\frac{\partial \tilde{\varphi}}{\partial \theta}\right)_{r=\mathcal{R}^-} = 0. \qquad (3)$$

Here, $\mu$ and $\mu_a$ are, respectively, diagonal and off-diagonal components of the permeability tensor $\ddot{\mu}$. One can compare Eq. (3) with Eq. (1). In both cases, there are the terms with the first-order derivative of the wave function with respect to the azimuth coordinate. In a case of the MS-potential wave solutions, one can distinguish the time direction (given by the direction of the magnetization precession and correlated with a sign of $\mu_a$) and the azimuth rotation direction (given by a sign of $\frac{\partial \tilde{\varphi}}{\partial \theta}$). For a given sign of a parameter $\mu_a$, there are different MS-potential wave functions, $\tilde{\varphi}^{(+)}$ and $\tilde{\varphi}^{(-)}$, corresponding to the positive and negative directions of the phase variations with respect to a given direction of azimuth coordinates, when $0 \leq \theta \leq 2\pi$. So a function $\tilde{\varphi}$ is not a single-valued function. It changes a sign when $\theta$ is rotated by $2\pi$ [6].

Similar to the Sagnac effect in optical microcavities, MDM oscillations in a quasi-2D ferrite disk are described by the Bessel-function azimuthally rotating waves. There are microwave-frequency rotating field configurations with power-flow vortices. Such fields with the azimuthally unidirectional energy transport around the disk axis are observed both inside and outside a ferrite sample. For a given direction of a bias magnetic field, the power-flow vortices are the same in the vacuum near-field region above and below the disk. Fig. 1 shows schematically such power flow rotations. The fact of the presence of the power flow circulations in the planes inside the ferrite and in vacuum above and below the disk arises a question on mechanical stability of the ferrite sample. Really, the disk body should sense the impact of the torque due to power-flow circulations and so should rotate mechanically. However, in all of the studies of MDM oscillations in a ferrite disk [6 – 1, 13 – 15], the entire microwave structure with an embedded ferrite sample, is mechanically fixed.

The field orbital rotation (and the power-flow circulations related to this field rotation) are due to topological effects in MDM oscillations. The fields are orbitally driven to satisfy properly the boundary conditions on a surface of a ferrite disk. Because of path-dependent interference, we observe topological-phase effects in the near-field region of the ferrite disk [6 – 11]. Nevertheless, the entire system (the microwave structure with an embedded ferrite disk) should be integrated and so, the system should be under the angular-momentum-balance conditions. In other words, there should exist conditions for neutrality of the vortex topological charges. Such the angular momentum balance can be realized only when the *opposite power flow circulations* are created on metal walls of the microwave structure, in which the ferrite disk is embedded (see Fig. 2). It means that at the MDM resonances in a ferrite disk, specific (topologically distingushed) conductivity electric currents should be induced on the metal parts of a microwave structure. The entire microwave structure, external to a ferrite disk, should "work out" the situation to give the possibility for integration of the entire Maxwell-equation system. The recoil from the metal parts of the microwave structure is not similar to the classical Lenz's law. It takes place via topological effects on a metal surface: creation of effective charges and surface chiral currents and recoil ME fields. There is the "topological recoil". The coupling between an electrically neutral MDM ferrite disk and an electrically neutral metal objects should involve the electromagnetic-induction and Coulomb interaction through virtual photons mediation. Moreover, there is the discrete-state process related to the quantization properties of MDM oscillations, in which quantized virtual field excitations appear in vacuum.

Why there is a topological nature of the rotating fields in a MDM ferrite disk? A ferrite is a magnetic dielectric with low losses. This may allow for electromagnetic waves to penetrate the ferrite and results in an effective interaction between the electromagnetic waves and magnetization within the ferrite. Such an interaction can demonstrate interesting physical features. Because of gyrotropy, electromagnetic waves incident on a ferrite-dielectric interface have reflection symmetry breaking. In Ref. [19], different cases of reflection of electromagnetic waves from a ferrite-dielectric interface were studied. It was shown that in a certain range of the material parameters, a frequency region, and a region of a bias magnetic field, the phase shift on total reflection of electromagnetic waves from a lossless ferrite is nonreciprocal and magnetically tunable. The reflection of microwaves at a dielectric-ferrite interface from the view point of ray and energy propagation shows [20, 21] that for a given direction of a bias magnetic field one has unidirectional energy transport along the interface. Thus, on a lateral surface of a normally magnetized ferrite disk, one has an azimuthally nonreciprocal phase shift of electromagnetic waves reflected from a ferrite (see Fig.3). The MDM oscillations in a ferrite-disk sample give evidence for quantized effects of interaction between the electromagnetic waves and magnetization within the ferrite. In this case a nonreciprocal phase at the orbit circulation on a lateral surface is quantizes as well.

In the present paper, we analyze numerically the interaction of an MDM ferrite disk with a metal screen. We show that to the topological action of the azimuthally unidirectional transport of energy in a MDM-resonance ferrite sample there exists the opposite topological reaction on a metal screen placed near this sample. We call this effect topological Lenz's effect. The topological Lenz's law is applied to opposite topological charges: one in a ferrite sample and another on a metal screen. The MDM-originated near fields – the magnetoelectric (ME) fields – induce helical surface electric currents and effective charges on a metal. The ME fields formed by these currents and charges will oppose their cause.

## II. RESULTS

Compared to our previous numerical studies [7, 9 – 11, 22, 23] of MDM oscillations in a ferrite disk placed in standard microwave rectangular waveguides, in this work we use extremely thin rectangular waveguide. In such a structure, the waveguide metal walls are situated very close to the ferrite-disk surface. We also study a microwave microstrip structure with an embedded MDM ferrite disk. In the structures under consideration we are able to consider the angular-momentum balance conditions separately from the balance for a linear momentum of propagating waves.

In the analysis, we use the same disk parameters as in Refs. [7, 9 – 11, 22, 23]: The yttrium iron garnet (YIG) disk has a diameter of $2\mathcal{R} = 3$ mm and the disk thickness is $t = 0.05$ mm; the disk is normally magnetized by a bias magnetic field $H_0 = 4900$ Oe; the saturation magnetization of the ferrite is $4\pi M_s = 1880$ G. For better understanding the field structures, we assume in our numerical studies that a ferrite disk has very small losses: The linewidth of a ferrite is $\Delta H = 0.1$ Oe. A ferrite disk is placed inside a thin $TE_{10}$-mode rectangular waveguide symmetrically to its walls so that the disk axis is perpendicular to a wide wall of a waveguide. The waveguide walls are made of a perfect electric conductor (PEC). A rectangular waveguide cross-section sizes are $a = 22.86$ mm, $b = 1$ mm. Fig. 4 shows a thin rectangular waveguide with an embedded quasi-2D ferrite disk. The entire microwave structure (a ferrite disk and a waveguide) can be considered as a quasi-2D structure.

Figure 5 shows the module of the reflection and transmission coefficients (the $S_{11}$ and $S_{21}$ scattering-matrix parameters, respectively). The resonance peaks are designated in accordance with the mode classification used in Ref. [17]. This classification shows the number of variations of the MS-potential function in a ferrite disk with respect to azimuth and radial coordinates. In the present analysis we use the 1st and 2nd radial modes, i. e. the modes with one and two radial variations of the MS-potential function, respectively. For both these modes, the azimuth number is equal to 1. Fig. 6 shows the Poynting vector distributions for the 1st radial mode on the upper plane of a ferrite disk for two directions of the electromagnetic wave propagation in a rectangular waveguide and at the same direction of a bias magnetic field. In a view along $-z$ direction, one clearly sees the CCW power-flow vortex. We will classify this vortex by the topological charge $Q = +1$. On a metal screen (the upper wall of a waveguide), we have the CW power-flow vortex, which can be classified by the topological charge $Q = -1$ (see Fig. 7). The view is along $-z$ direction on an upper waveguide wall from the outside metal region. A vacuum gap between the upper plane of a ferrite disk and the upper wall of a waveguide is sufficiently narrow. On vacuum planes inside such a gap, we can see the effect of counteraction of the two topological charges. On a certain vacuum plane inside this gap, the condition of "topological neutrality" should exist. The power-flow distribution on a vacuum plane 20 um above the upper plane of a ferrite disk in Fig. 8, shows that the vortex structure, originated from a ferrite disk, becomes much emasculated.

The fields in vacuum near a MDM ferrite disk – the ME fields – are determined by the magnetization distributions. The vector field of RF magnetization $\vec{m}$ in a ferrite disk has two parts: the potential ($\vec{\nabla} \cdot \vec{m} \neq 0$) and curl ($\vec{\nabla} \times \vec{m} \neq 0$) ones. While the magnetic component of the ME field is originated from the potential part of the magnetization, the electric component of the ME field is originated from the curl part of the magnetization [10]. An analytical expression for magnetization $\vec{m}$ is given in Ref. [24]. For the known disk parameter, we calculated analytically the magnetization for the 1st radial mode. The distribution shown in Fig. 9 gives evidence for the spin and orbital angular momentums of the magnetization field. In the orbitally driven distribution of $\vec{m}$, one can clearly observe the regions of magnetization vortices. At the same time, it is evident that the fields in vacuum near a metal wall are related to the electric current distributions on a metal surface. Fig. 10 shows the orbitally driven electric current distributions on the upper wall of a waveguide for the 1st radial mode. In this case, there exist the regions of current vortices. For the present study, it is worth noting that the in-plane rotating distributions in Figs. 9 and 10 are very similar topologically one to another.

The structures of the fields of a MDM ferrite disk were studied numerically and analytically in Refs. [7, 9 – 11, 22, 23, 25]. The Poynting-vector vortices of such fields on a disk plane are formed by the magnetic-field components normal to the disk surface and the electric-field components tangential to the disk surface: $\vec{E}_t^{(ferrite)} \times \vec{H}_n^{(ferrite)}$. Because of the presence of surface electric charges and electric currents induced on a closely situated metal wall by the fields originated from MDM oscillations, the Poynting-vector vortices of on a screen are formed by the electric-field components normal to the metal surface and the magnetic-field components tangential to the metal surface: $\vec{E}_n^{(metal)} \times \vec{H}_t^{(metal)}$. The field structures shown in Fig.11, can explain qualitatively why the power-flow circulations on a plane of a ferrite disk and on a metal screen are oppositely directed.

The topological nature of the observed effect becomes clear when one analyzes more in details the electric and magnetic fields on a metal wall. Fig. 12 shows the magnetic field distribution for the 1st radial mode on the upper wall of a waveguide. One can see specific regions of surface topological magnetic charges (STMCs). Similar STMCs were observed in previous studies when a MDM ferrite disk was placed inside a standard ("thick") waveguide [9, 22]. The STMCs are points of divergence and convergence of a 2D magnetic field (or a surface magnetic flux density $\vec{B}_S$) on a waveguide wall. As is evident from Fig. 12, one has nonzero outward (inward) flows of the vector field $\vec{B}_S$ through a closed flat loop $\mathcal{L}$ lying on the metal wall and surrounding the points of divergence or convergence: $\oint_{\mathcal{L}} \vec{B}_S \cdot \vec{n}_S d\mathcal{L} \neq 0$. Here $\vec{n}_S$ is a normal vector to contour $\mathcal{L}$, lying on a metal surface. At the same time, it is clear, however, that $\vec{\nabla}_S \cdot \vec{B}_S = 0$, since there are zero magnetic fields at the points of divergence or convergence. Such topological singularities on the metal waveguide wall show unusual properties. One can see that, for the region bounded by the circle $\mathcal{L}$, no planar variant of the divergence theorem takes place. It is worth noting also that the STMCs are observed exactly at the same places on a metal wall where the centers of the electric current vortices are situated (see Fig. 10).

Distributions of the magnetic- and electric-field components, on the metal surface, are also orbitally driven with the same angular velocity as other distributions (such as magnetization in a ferrite and surface magnetic current on a wall). Importantly, both these, electric and magnetic, fields have their maximums (minimums) exactly at the same azimuth coordinates. This result in appearance very interesting topological structures on a metal wall originated from the MDM resonances in the disk. The fields on an upper metal wall for the 1st radial mode, shown in Fig. 13, give evidence for clearly distinguished regions of the surface electric and topical-magnetic charges. The "color" picture shows a distribution of a normal component of the electric field $\vec{E}_n^{(metal)}$ while the "arrow" picture shows a distribution of a tangential component of the magnetic field field $\vec{H}_t^{(metal)}$. The time phase of $\omega t = 255°$ corresponds to the case when the electric and topical-magnetic dipoles are perpendicular to the waveguide axis. The distributions in Fig. 13 show that on a metal wall there is a "glued pair" of 2D rotating dipoles: the electric and topological magnetic ones. To a certain extent (in our case, there is a 2D, topological, and rotating structure) we have an analog of a Tellegen particle [26, 27].

On a surface of a MDM ferrite disk, there are both the regions of the orbitally driven normal magnetic field $\vec{B}_n$ and normal electric field $\vec{E}_n$. Importantly, the maximums (minimums) of the rotating fields $\vec{E}_n$ and $\vec{B}_n$ are situated at the same places on a disk plane [7, 10, 25]. This is illustrated in Fig. 14 for a certain time phase $\omega t$. This field structure is projected on a metal wall closely placed to a ferrite disk. Together with surface electric charges $\rho_s$ induced on a metal wall by the electric field $\vec{E}_n$, there also the Faraday-law eddy currents induced on a metal surface by the time-derivative of a normal component of the rotating MDM magnetic field $\vec{B}_n$. Evidently, the induced surface electric current $\vec{j}_s$ should have two components. There are the linear currents arising from the continuity equation for surface electric charges, for which $\vec{\nabla}_S \cdot \vec{j}_S \neq 0$, and the Faraday-law eddy currents, for which $\vec{\nabla}_S \times \vec{j}_S \neq 0$. As a result, we have the sources of the linear-current and eddy-current components situated at the same place on a metal wall. Thus, the form of a surface electric current is not a closed line. It should be a spiral. This form of a surface electric current can be traced on the pictures in Fig. 10. More clearly, the current lines, modeled by the spirals, are shown in Fig. 15. There are both the right-handed and left-handed flat spirals.

The effect of the angular-momentum opposite topological reaction on a metal screen for the 2nd radial mode in a "thin" waveguide with a ferrite sample is shown in Ref. [24]. Also, the angular-momentum-balance condition has been proven in a microwave microstrip structure with an embedded MDM ferrite disk [24].

### III.     DISCUSSION AND CONCLUSION

Long-range magnetic-dipolar interactions in confined magnetic structures are not in the scope of classical electromagnetic problems and, at the same time, have properties essentially different from the effects of exchange ferromagnetism. The MDM spectral properties in confined magnetic structures are based on postulates about physical meaning of the magnetostatic (MS) potential function $\psi(\vec{r},t)$ as a complex scalar wave function, which presumes long-range (on the scales much bigger than the exchange-interaction scales) phase coherence. The MDMs in a quasi-2D ferrite disk are characterized by energy-eigenstate orthogonality relations. Due to topological states on a lateral surface of a ferrite disk, these modes have orbital angular momentums. In a vacuum subwavelength region abutting to a MDM ferrite disk, one can observe the quantized-state power-flow vortices. In such a region, a coupling between the time-varying electric and magnetic fields is different from such a coupling in regular electromagnetic fields. These specific near fields, originated from MDM oscillations, we term magnetoelectric (ME) fields. The ME field solutions give evidence for spontaneous symmetry breaking at the resonant states of MDM oscillations [10]. As a source of the ME field, there is the pseudoscalar parameter of magnetization helicity. This parameter, appearing since the magnetization $\vec{m}$ in a ferrite disk has two parts: the potential and curl ones, is calculated as

$$V \equiv \text{Im}\left[\vec{m}\cdot\left(\vec{\nabla}\times\vec{m}\right)^*\right] = \frac{1}{\omega c}\text{Re}\left[\vec{j}^{(m)}\cdot\left(\vec{j}^{(e)}\right)^*\right], \qquad (4)$$

where $\vec{j}^{(m)} = i\omega\vec{m}$ and $\vec{j}^{(e)} = c\vec{\nabla}\times\vec{m}$ are, respectively, the magnetic and electric current densities in a ferrite medium, $c$ is the light velocity in vacuum [10, 27, 28]. For the MS-potential wave function in a ferrite disk represented by Eq. (2), the pseudoscalar parameter $V$ was derived analytically in [24]. This parameter, analytically calculated for the 1st radial mode (with normalization of the mode amplitude coefficient $C$ [25]), is shown in Fig. 16. The pseudoscalar parameter $V$ gives evidence for the presence of two coupled and mutually parallel currents – the electric and magnetic ones – in a localized region of a microwave structure. This is the cause of appearance of flat-spiral currents and a "glued pair" of 2D rotating dipoles on a metal wall.

In a microwave structure with an embedded ferrite disk, an orbital angular momentum, related to the power-flow circulation, must be conserved in the process. Thus, if power-flow circulation is pushed in one direction in a ferrite disk, then the power-flow circulation on metal walls to be pushed in the other direction by the same torque at the same time. In the present paper, we analyzed numerically the interaction of an MDM ferrite disk with a metal screen. Our numerical studies give evidence for the fact that to the topological action of the azimuthally unidirectional transport of energy in a MDM-resonance ferrite sample there exists the opposite topological reaction on a metal screen placed near this sample. We call this effect topological Lenz's effect. The topological Lenz's law is applied to opposite topological charges: one in a ferrite sample and another on a metal screen. The MDM-originated near fields – the magnetoelectric (ME) fields – induce helical surface electric currents and effective charges on a metal. The ME fields formed by these currents and charges will oppose their cause. As we have shown, the in-plane magnetization distribution in a ferrite disk and the surface current distribution on a metal wall are very similar topologically one to another. These ME sources, in a ferrite and on a metal wall, result in appearance of the ME field in the vacuum gaps. To characterize the ME properties of the field, we use distribution of the normalized helicity factor in a vacuum region. The normalized helicity factor, calculated as

$$\cos\alpha = \frac{\text{Re}\left(\vec{E}\cdot\vec{H}^*\right)}{\left|\vec{E}\right|\left|\vec{H}\right|}, \tag{5}$$

gives evidence to the fact that the electric and magnetic fields in vacuum are not mutually perpendicular [10, 22, 23, 25]. The normalized helicity factor distribution on the *xz* cross-sectional plane of the vacuum gap, numerically calculated for the fields of the 1st radial mode, is shown in Fig. 17. One can see that a maximal angle between the electric and magnetic fields in vacuum is about $70°$.

In the present study, we used microwave waveguiding structures with the metal walls situated very close to the ferrite-disk surfaces. The vacuum gaps between the metal and ferrite are much less than a diameter of a MDM ferrite disk and thus the entire microwave structure (a ferrite disk and a waveguide) can be considered as a quasi-2D structure. In the shown distributions, the regions of the power-flow circulations in a ferrite disk and on metal wall are, in fact, opposite one another and the angular-momentum balance does not depend, actually, on the direction of the electromagnetic wave propagation in a waveguide. In this case, we have exchange of two types of virtual photons: (*a*) the static electric force and (*b*) the electromagnetic induction. The observed properties rely on ME virtual photons to act as the mediator.

However, the fields in a ferrite disk rotate at microwave frequencies and situation becomes more complicated when the vacuum region scale is about the disk diameter or more and the finite speed of wave propagation in vacuum – the retardation effects – should be taken into consideration. This means that for a brief period the total angular momentum of the two topological charges (one in a ferrite, another on a metal) is not conserved, implying that the difference should be accounted for by an angular momentum in the fields in the vacuum space. The magnetization dynamics have an impact on the phenomena connected with fluctuation energy in vacuum. Because of MDM resonances, this angular momentum is quantized. As the rotational symmetry is broken in this case, the Casimir torque [29 – 33] arises because the Casimir energy now depends on the angle between the directions of the magnetization vectors in a ferrite and electric-current vectors on a metal wall. A vacuum-induced Casimir torque allows for torque transmission between the ferrite disk and metal wall avoiding any direct contact between them.

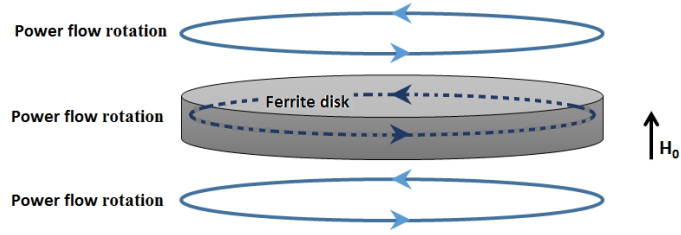

Fig. 1. Power flow rotations. For a given direction of a bias magnetic field, the power-flow circulations are the same inside a ferrite and in the vacuum near-field regions above and below the disk.

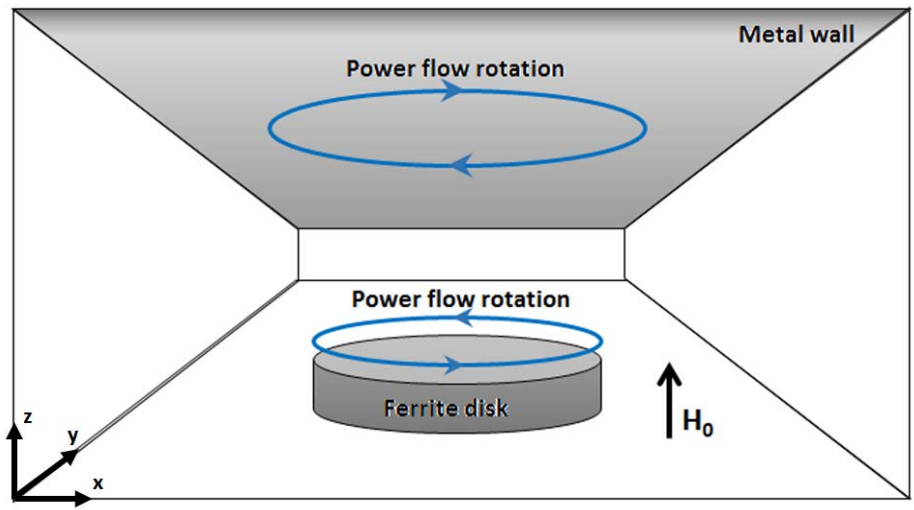

Fig. 2. Angular momentum balance conditions: In a view along *z* axis, there are opposite power flow circulations on a surface of a ferrite disk and on a metal surface. At the same time, in a view along *y* axis, both power flow circulations are clockwise rotations.

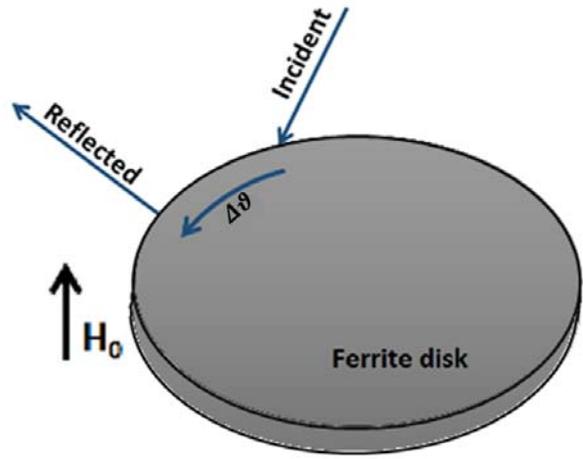

Fig. 3. The topological nature of the rotating fields in a MDM ferrite disk. On a lateral surface of a normally magnetized ferrite disk, one has an azimuthally nonreciprocal phase shift of electromagnetic waves reflected from a ferrite.

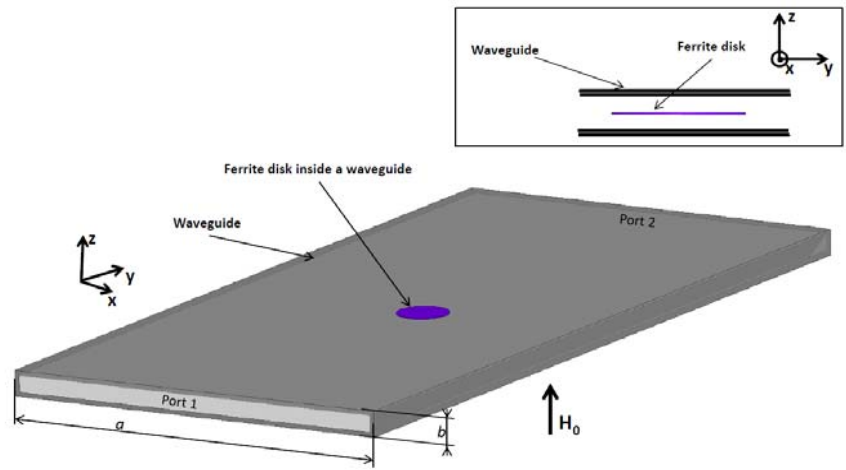

Fig. 4. A thin rectangular waveguide with an embedded quasi-2D ferrite disk. An insert shows the disk position inside a waveguide.

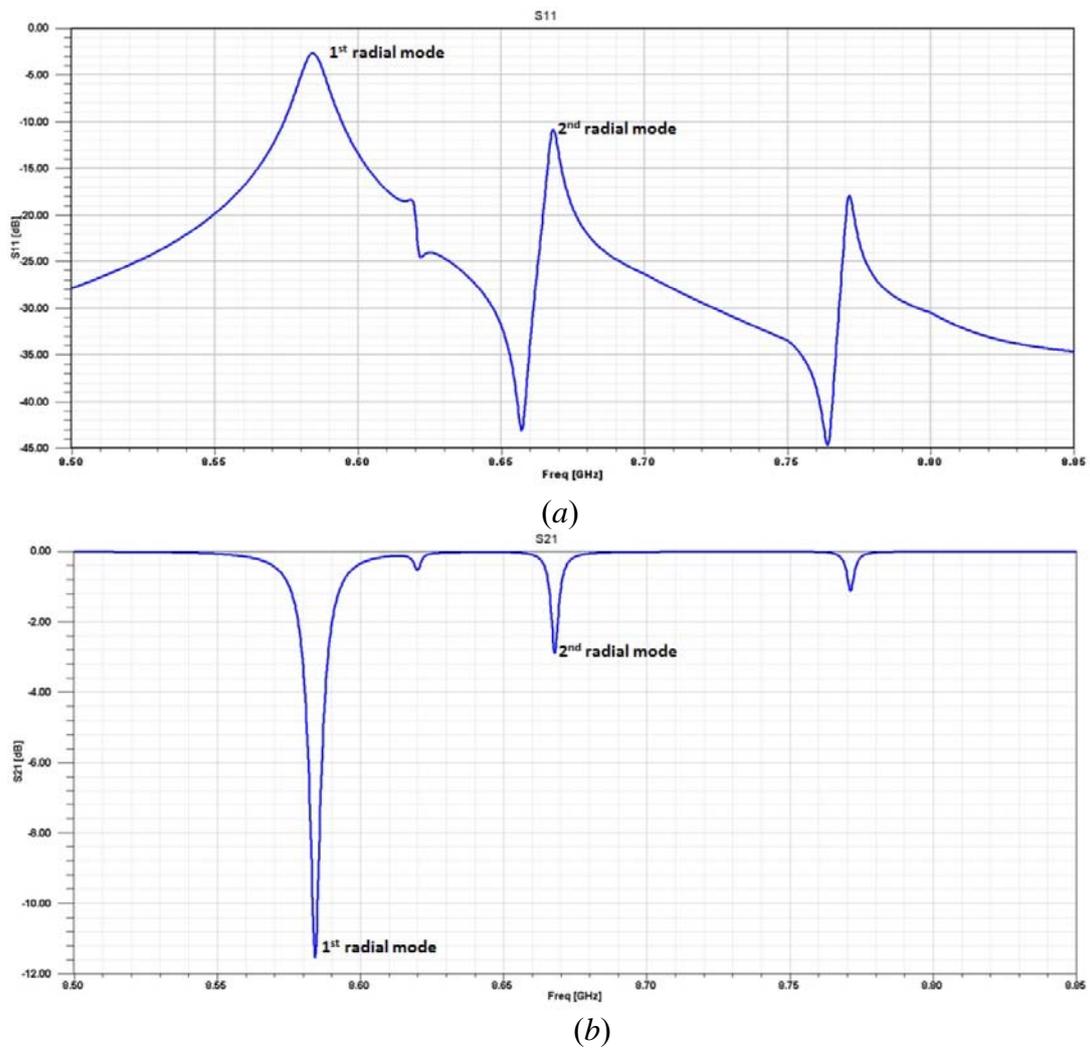

Fig. 5. The MDM spectra in a waveguide. (*a*) Reflection coefficient; (*b*) transmission coefficient.

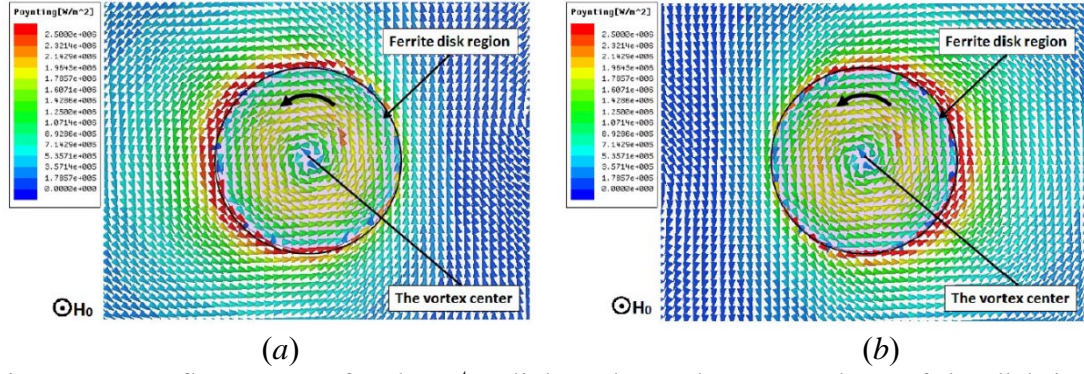

Fig. 6. Power-flow vortex for the 1st radial mode on the upper plane of the disk in a view along $-z$ direction. (*a*) Electromagnetic wave propagation in a waveguide from port 1 to port 2 (1 ➔ 2), (*b*) wave propagation from port 2 to port 1 (2 ➔1). A black arrow clarifies the power-flow direction. A bias magnetic field is directed along $+z$ direction. The vortex is classified by the topological charge $Q = +1$.

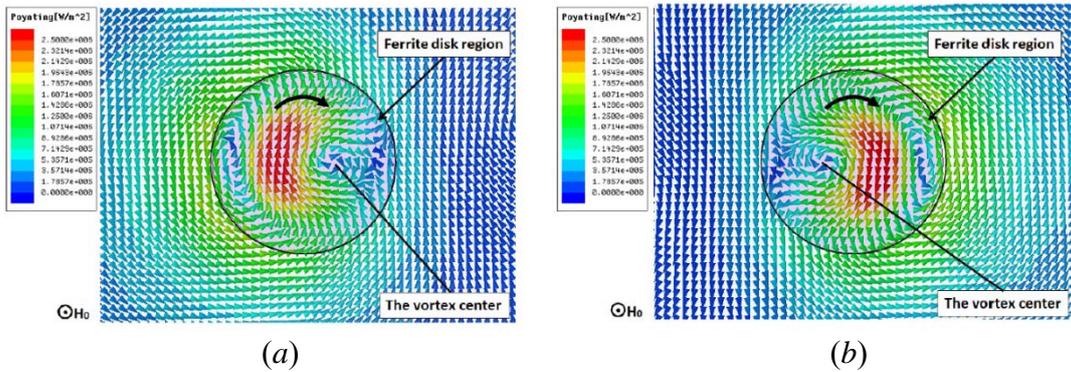

Fig. 7. Power-flow vortex for the 1st radial mode on the upper wall of a waveguide. (*a*) Wave propagation 1 ➔ 2, (*b*) wave propagation 2 ➔1. A black arrow clarifies the power-flow direction. A bias magnetic field is directed along $+z$ direction. The view is along $-z$ direction on an upper waveguide wall from the outside metal region. The vortex is classified by the topological charge $Q = -1$.

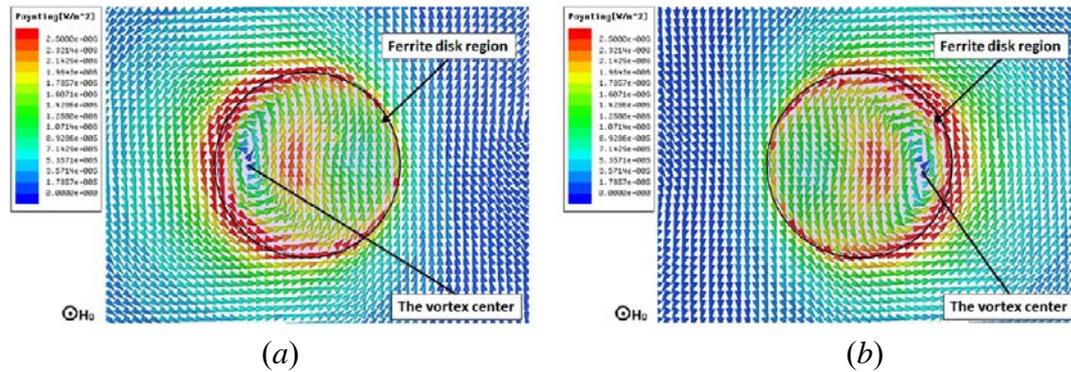

Fig. 8. Power-flow distribution 1st radial mode on a vacuum plane 20 um above the upper plane of a ferrite disk. (*a*) Wave propagation 1 ➔ 2, (*b*) wave propagation 2 ➔1. A bias magnetic field is directed along $+z$ direction. The view is along $-z$ direction. The vortex structure with the charge $Q = +1$ is much emasculated.

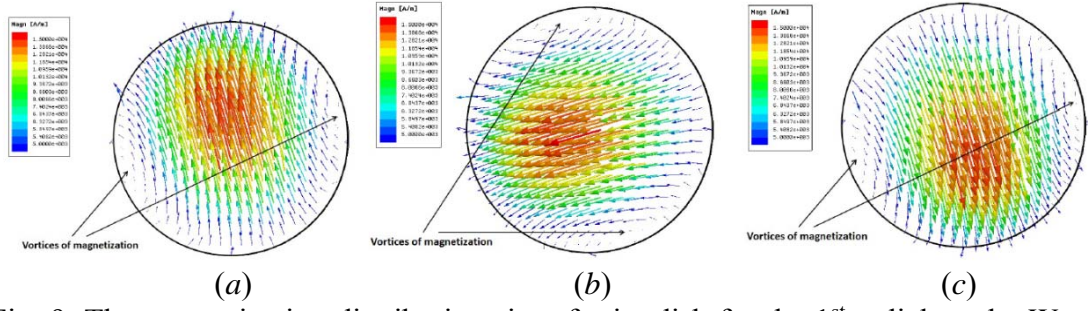

(a)                        (b)                        (c)

Fig. 9. The magnetization distributions in a ferrite disk for the 1$^{st}$ radial mode. Wave propagation 1 ➔ 2. A bias magnetic field is directed along $+z$ direction. (a) Time phase $\omega t = 0°$, (b) $\omega t = 90°$, (c) $\omega t = 180°$. The regions of magnetization vortices are clearly observed.

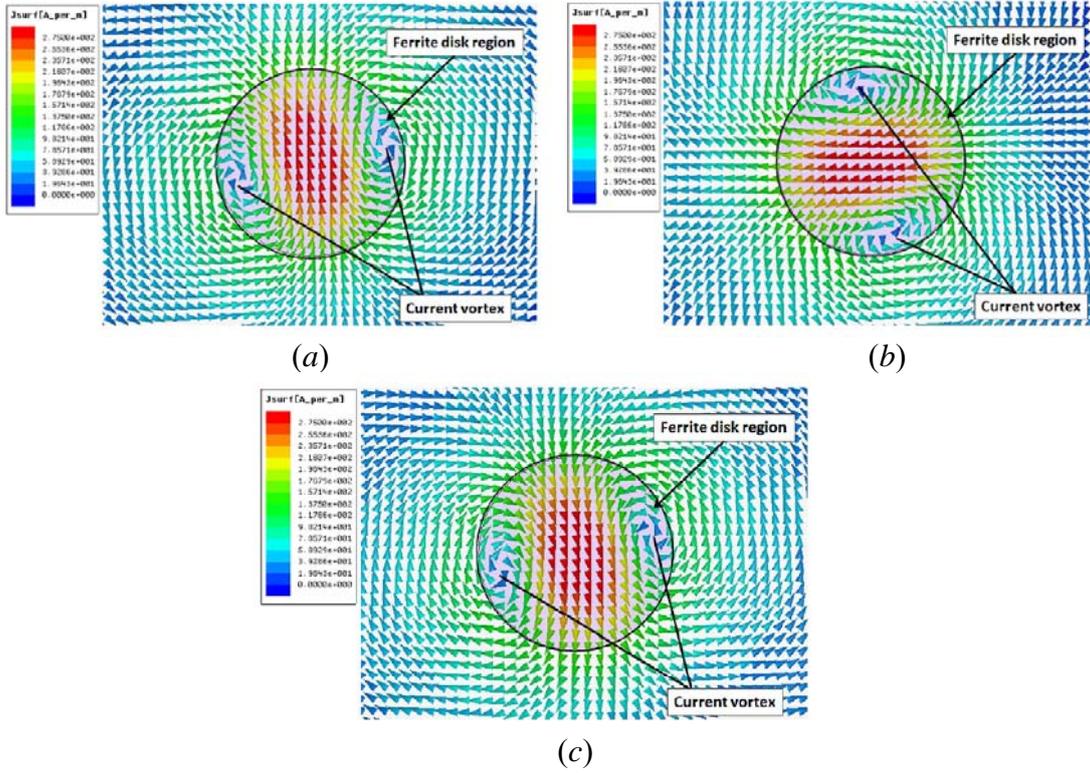

(a)                                   (b)

(c)

Fig. 10. Surface electric current for the 1$^{st}$ radial mode on the upper wall of a waveguide. Wave propagation 1 ➔ 2. A bias magnetic field is directed along $+z$ direction. (a) Time phase $\omega t = 0°$, (b) $\omega t = 90°$, (c) $\omega t = 180°$. One can see the current vortices.

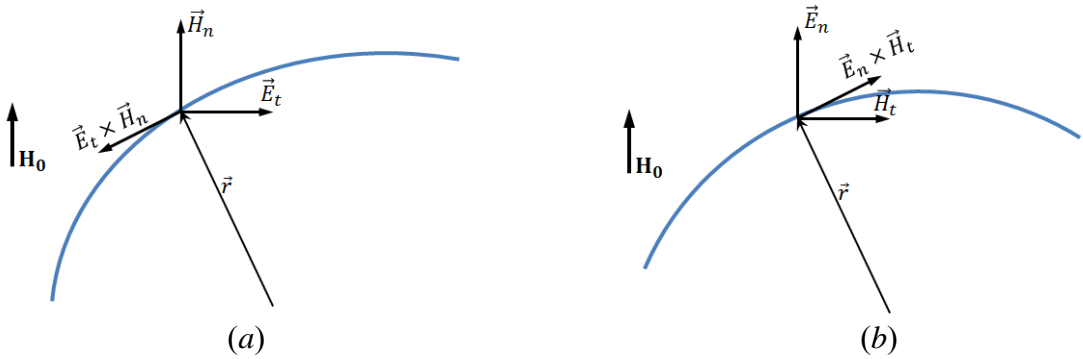

(a)                                   (b)

Fig 11. The fields on lines of the power flow circulations. (*a*) On the ferrite-disk plane, (*b*) on the metal-wall plane. The angular-momentum vectors $\vec{r} \times \left( \vec{E}_t^{(\text{ferrite})} \times \vec{H}_n^{(\text{ferrite})} \right)$ and $\vec{r} \times \left( \vec{E}_n^{(\text{metal})} \times \vec{H}_t^{(\text{metal})} \right)$ are oppositely directed.

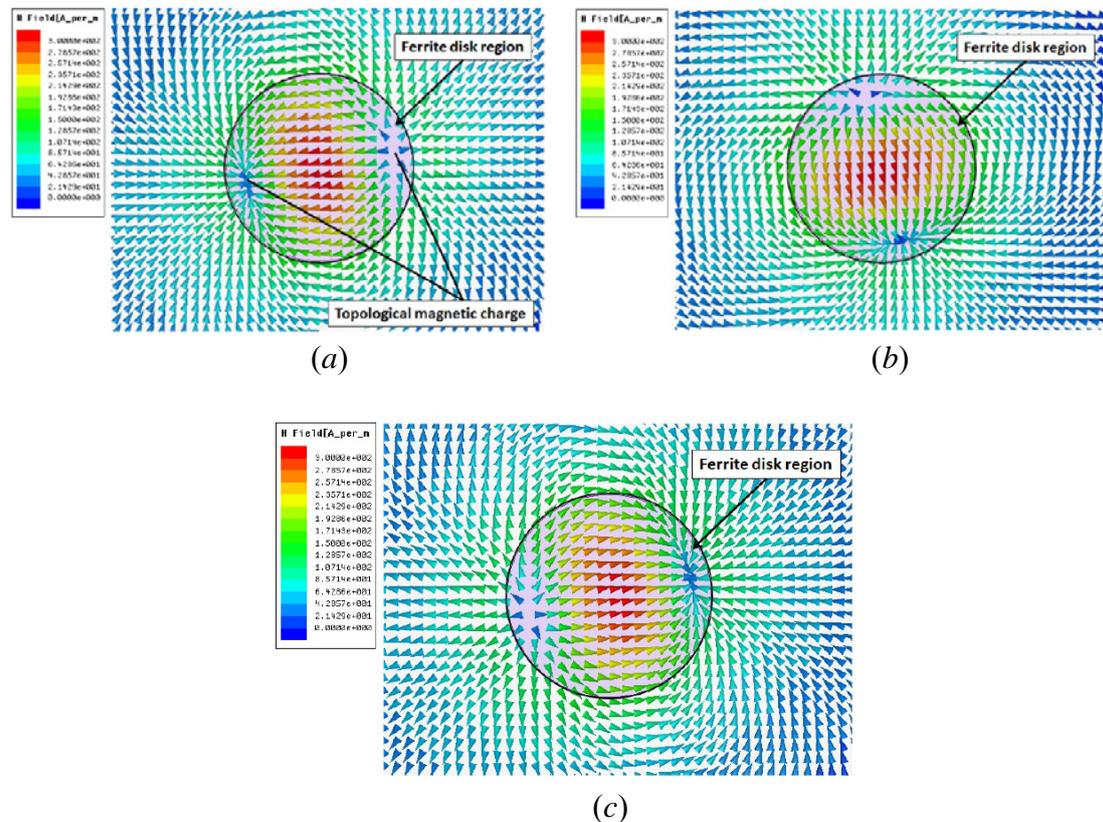

Fig. 12. Magnetic field distribution for the 1$^{\text{st}}$ radial mode on the upper wall of a waveguide. Wave propagation 1 ➔ 2. A bias magnetic field is directed along $+z$ direction. (*a*) Time phase $\omega t = 0°$, (*b*) $\omega t = 90°$, (*c*) $\omega t = 180°$.

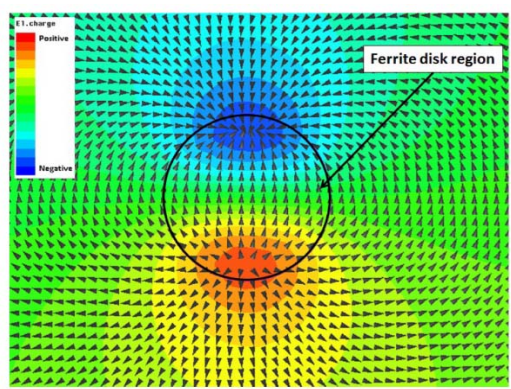

Fig. 13. The orbitally driven fields on an upper metal wall for the 1$^{\text{st}}$ radial mode. The "color" picture shows a distribution of a normal component of the electric field $\vec{E}_n^{(\text{metal})}$ while the "arrow" picture shows a distribution of a tangential component of the magnetic field field $\vec{H}_t^{(\text{metal})}$. The regions of the surface electric and topical-magnetic

charges are clearly distinguished. The time phase of $\omega t = 255°$ corresponds to the case when the electric and topical-magnetic dipoles are perpendicular to the waveguide axis.

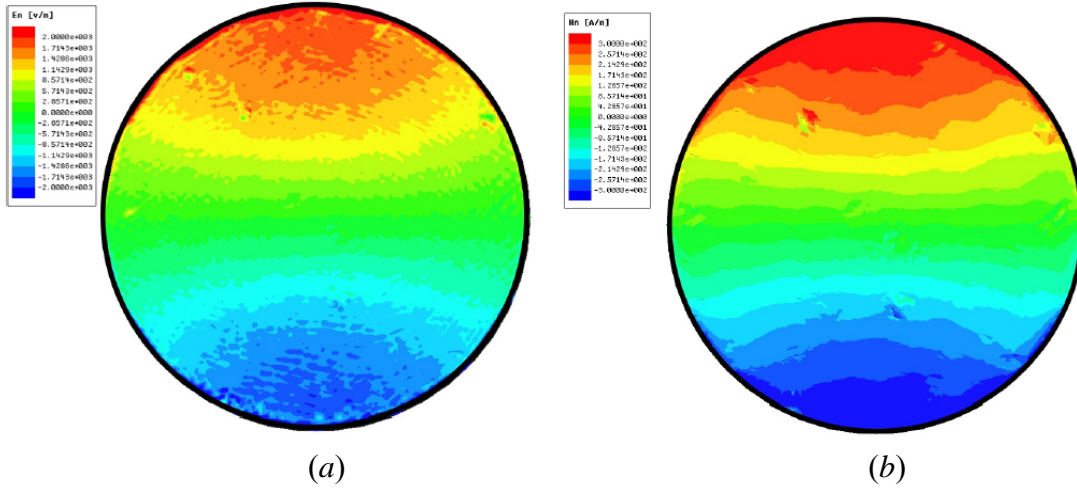

(a)  (b)

Fig. 14. The magnitudes of the normal electric field $\vec{E}_n$ (a) and normal magnetic field $\vec{B}_n$ (b) on a surface of a MDM ferrite disk at a certain time phase $\omega t$. The fields are orbitally driven and mutually synchronized. The maximums (minimums) of the rotating fields $\vec{E}_n$ and $\vec{B}_n$ are situated at the same places on a disk plane.

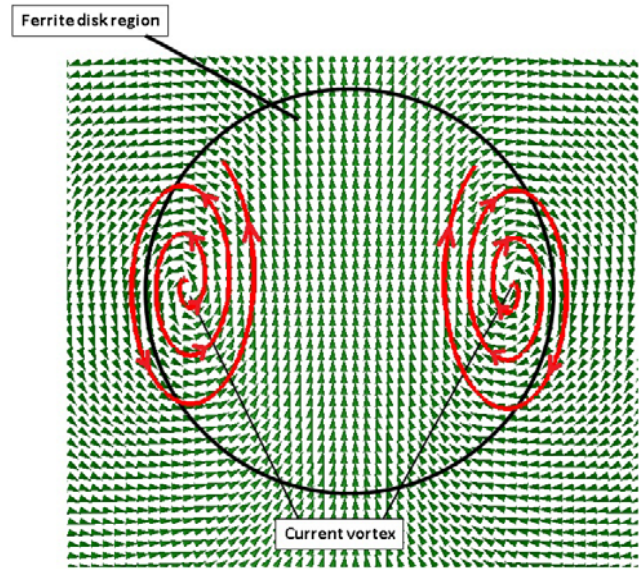

Fig. 15. Surface electric current for the 1$^{st}$ radial mode on the upper wall of a waveguide shown schematically as the right-handed and left-handed flat spirals (red lines).

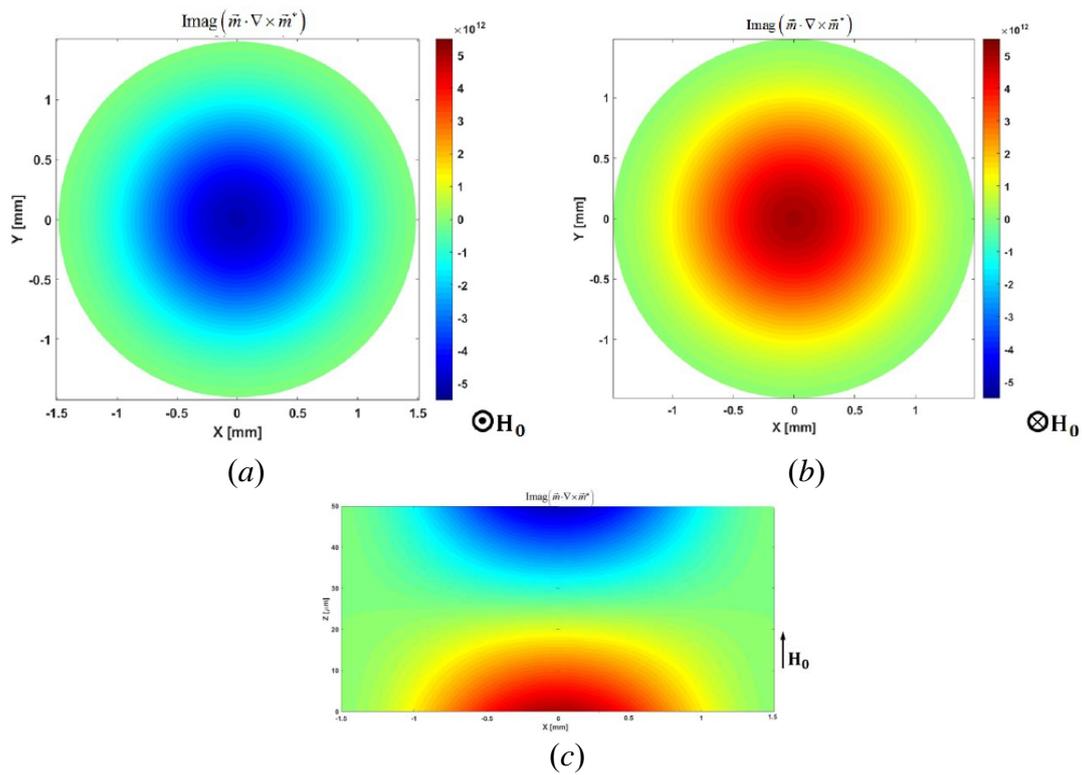

Fig. 16. Pseudoscalar parameter $V = \text{Im}\left[\vec{m} \cdot (\nabla \times \vec{m})^*\right]$ on the upper plane of a ferrite disk theoretically calculated for the 1st radial mode. (*a*) On the upper plane of a ferrite disk at a bias field directed along *z* axis; (*b*) on the upper plane of a ferrite disk at a bias field directed along -*z* axis; (c) on the *xz* cross-sectional plane of a ferrite disk at a bias field directed along *z* axis.

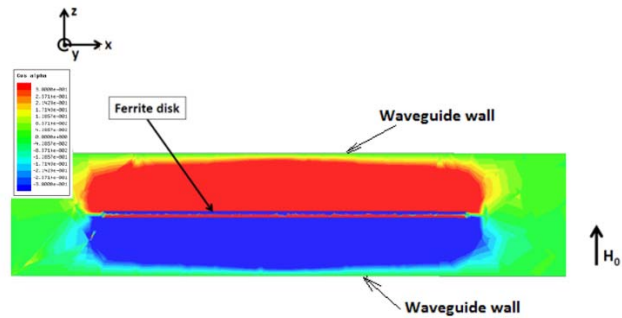

Fig. 17. The normalized helicity factor for the fields on the *xz* cross-sectional plane of the vacuum gaps.

# Supplementary information for "Azimuthally unidirectional transport of energy in magnetoelectric fields: Topological Lenz's effect"

R. Joffe [1,2], E. O. Kamenetskii [1], and R. Shavit [1]

[1] Microwave Magnetic Laboratory, Department of Electrical and Computer Engineering, Ben Gurion University of the Negev, Beer Sheva, Israel

[2] Department of Electrical and Electronics Engineering, Shamoon College of Engineering, Beer Sheva, Israel

January 17, 2017

In the supplementary material we show the effect of the angular-momentum opposite topological reaction on a metal screen for the 2nd radial MDM in a "thin" waveguide with a ferrite sample. We prove numerically the angular-momentum-balance condition in a microwave microstrip structure with an embedded MDM ferrite disk. Also, we derive and calculate analytically a pseudoscalar parameter of magnetization helicity.

### A. The 2nd radial MDM in a "thin" waveguide with a ferrite disk

Power-flow vortices in a ferrite disk are observed for every MDM in the spectrum and so there should exist neutrality for the vortex topological charges for all the modes. We illustrate here the angular momentum balance for the 2nd radial MDM in a "thin" waveguide with a ferrite sample. Fig. 1 shows the power-flow vortices for the 2nd radial mode on the upper plane of the disk. One sees the two CCW and one CW power-flow vortices, which can be classified, successively, by three topological charges: $+1, -1, +1$. The angular momentum balance can be realized only when the opposite power flow circulations are created on metal parts of a microwave waveguide. Fig. 2 shows the power-flow vortices for the 2nd radial mode on the upper metal wall for two directions of the electromagnetic wave propagation in a rectangular waveguide and at the same direction of a bias magnetic field. The vortices are observed from the outside region above a waveguide along $-z$ direction. One sees that now there are two CW and one CCW power-flow vortices. In succession, the vortices are classified by the topological charges: $-1, +1, -1$. In a vacuum gap between the upper plane of a ferrite disk and the upper wall of a waveguide we have counteraction of the two topological charges. In Fig. 3, we can see that on a vacuum plane 20 um above the upper plane of a ferrite disk the vortex structure, originated from a ferrite disk, becomes emasculated. In comparison with the situation shown in Fig. 8 in the article, this emasculation is not so evident since the fields of the 2nd radial mode are concentrated more close to the disk plane and so the metal influence on this level in vacuum is not so strong.

### B. The 1st radial MDM in a microwave microstrip structure with

For more illustration of the angular momentum balance in microwave waveguides with an embedded MDM ferrite disk, we analyze also the situation with a microwave microstrip structure. The structure is shown in Fig. 4. There is the structure combined with two microstrip lines. The disk is placed between these lines on a surface of a dielectric substrate. The thickness of a dielectric substrate is 1.5 mm.

We consider the angular-momentum-balance conditions for the 1st radial MDM. Fig 5. shows the power-flow distribution on a bottom plane of the disk when a bias magnetic field is directed along $+z$ direction.. In a view along $-z$ direction, there is the CCW power-flow circulation. We classify this vortex by the topological charge $Q = +1$. In Fig. 6, one can see the power-flow distribution on a ground plane of a microstrip structure at the same direction of a bias magnetic field. In this case, in a view along $-z$ direction one observes the CW power-flow circulation. This vortex is classified by the topological charge $Q = -1$. Importantly, for a given direction of a bias magnetic field, the power-flow circulation on a metal region corresponding to the projection of the ferrite-disk area, is the same for two opposite directions of the wave propagation in the entire microwave structure.

Likewise to the surface electric current on the upper wall of a waveguide shown in the Article, the electric current distribution on a ground plane of a microstrip structure is orbitally driven and has the regions of current vortices. This current, shown in Fig. 7 for a certain time phase, is topologically similar to the magnetization distributions in a ferrite disk.

### C. Pseudoscalar parameter of magnetization helicity

MDM oscillations are described by a MS-potential wave function, which is related to the RF magnetic field as $\vec{H} = -\vec{\nabla}\psi$ [S1 – S3]. The MS-potential wave function in a ferrite disk with the disk axis oriented along $z$ axis, is written as: $\psi = C\xi(z)\tilde{\varphi}(r,\theta)$, where $\tilde{\varphi}$ is a dimensionless membrane function, $r$ and $\theta$ are in-plane coordinates of a cylindrical coordinate system, $\xi(z)$ is a dimensionless function of the MS-potential distribution along $z$ axis, and $C$ is a dimensional amplitude coefficient. The magnetization $\vec{m}$ is defined as $\vec{m} = -\vec{\tilde{\chi}} \cdot \vec{\nabla}\psi$, where $\vec{\tilde{\chi}}$ is the magnetic susceptibility tensor. For a ferrite disk magnetized along $z$ axis we have [S2, S3]

$$\vec{m} = -\begin{bmatrix} \chi & i\chi_a & 0 \\ -i\chi_a & \chi & 0 \\ 0 & 0 & 0 \end{bmatrix} \cdot \vec{\nabla}\psi. \quad (S1)$$

Assuming that MDMs are the fields rotating with an azimuth number $\nu$ $\left(\psi \propto e^{-i\nu\theta}\right)$, we obtain

$$\vec{m} = -\begin{bmatrix} \chi & i\chi_a & 0 \\ -i\chi_a & \chi & 0 \\ 0 & 0 & 0 \end{bmatrix} \cdot \vec{\nabla}\psi = C\xi(z)\left[\left(-\chi\frac{\partial\tilde{\varphi}}{\partial r} - \frac{\chi_a}{r}v\tilde{\varphi}\right)\hat{r} + i\left(\chi_a\frac{\partial\tilde{\varphi}}{\partial r} + \frac{\chi}{r}v\tilde{\varphi}\right)\hat{\theta}\right] \quad (S2)$$

The parameter of magnetization helicity is defined as the product $\vec{m}\cdot(\vec{\nabla}\times\vec{m})^*$. Since magnetization vector $\vec{m}$ has only in-plane components, we are interested in only in-plane components of the vector $\vec{\nabla}\times\vec{m}$ as well:

$$(\vec{\nabla}\times\vec{m})_{r,\theta} = -\frac{\partial m_\theta}{\partial z}\hat{r} + \frac{\partial m_r}{\partial z}\hat{\theta} = -C\frac{\partial\xi(z)}{\partial z}\left[i\left(\chi_a\frac{\partial\tilde{\varphi}}{\partial r} + \frac{\chi}{r}v\tilde{\varphi}\right)\hat{r} + \left(\chi\frac{\partial\tilde{\varphi}}{\partial r} + \frac{\chi_a}{r}v\tilde{\varphi}\right)\hat{\theta}\right]. \quad (S3)$$

As a result, we obtain

$$\vec{m}\cdot(\vec{\nabla}\times\vec{m})^* = -i2C^2\xi(z)\frac{\partial\xi(z)}{\partial z}\left(\chi\frac{\partial\tilde{\varphi}}{\partial r} + \frac{\chi_a}{r}v\tilde{\varphi}\right)\left(\frac{\chi}{r}v\tilde{\varphi} + \chi_a\frac{\partial\tilde{\varphi}}{\partial r}\right). \quad (S4)$$

We can see that the parameter of magnetization helicity is purely an imaginary quantity:

$$V \equiv \text{Im}\left[\vec{m}\cdot(\vec{\nabla}\times\vec{m})^*\right]. \quad (S5)$$

This parameter, appearing since the magnetization $\vec{m}$ in a ferrite disk has two parts: the potential and curl ones, can be also represented as

$$V = \frac{1}{\omega c}\text{Re}\left[\vec{j}^{(m)}\cdot(\vec{j}^{(e)})^*\right], \quad (S6)$$

where $\vec{j}^{(m)} = i\omega\vec{m}$ and $\vec{j}^{(e)} = c\vec{\nabla}\times\vec{m}$ are, respectively, the magnetic and electric current densities in a ferrite medium, $c$ is the light velocity in vacuum [S4 – S6]. This parameter, analytically calculated for the 1st radial mode, is shown in Fig. 16 of the article.

--------------------------

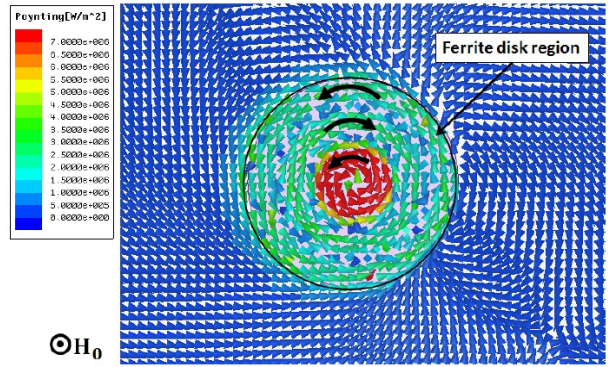

Fig. S1. Power-flow vortices for the 2nd radial mode on the upper plane of the disk. Electromagnetic wave propagation in a waveguide from port 1 to port 2 (1 ➔ 2). Black arrows clarify the power-flow directions. A bias magnetic field is directed along $+z$ direction. The vortices are classified by three topological charges. In succession, there are the topological charges $+1, -1, +1$.

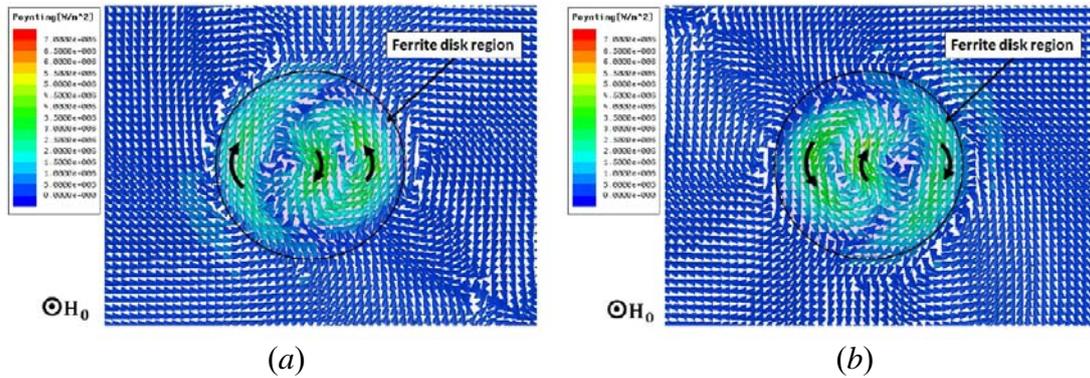

(*a*)                    (*b*)

Fig. S2. Power-flow vortices for the 2nd radial mode on the upper wall of a waveguide. (*a*) Wave propagation 1 ➔ 2, (*b*) wave propagation 2 ➔ 1. Black arrows clarify the power-flow directions. A bias magnetic field is directed along $+z$ direction. The view is along $-z$ direction on an upper waveguide wall from the outside metal region. The vortices are classified by three topological charges. In succession, there are the topological charges $-1, +1, -1$.

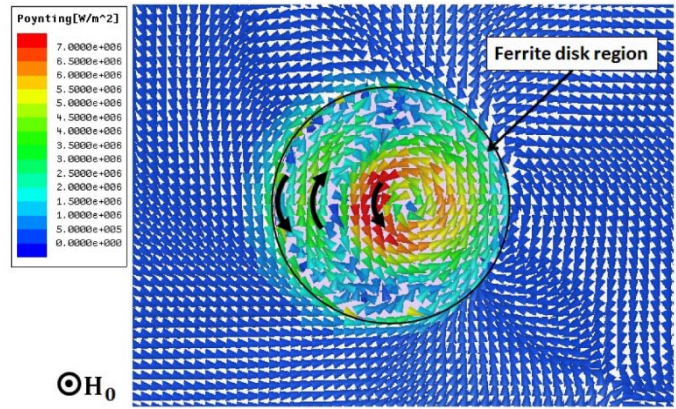

Fig. S3. The 2nd radial mode. 20 um above the upper plane of the disk. 1 ➔ 2. Black arrows clarify the power-flow directions.

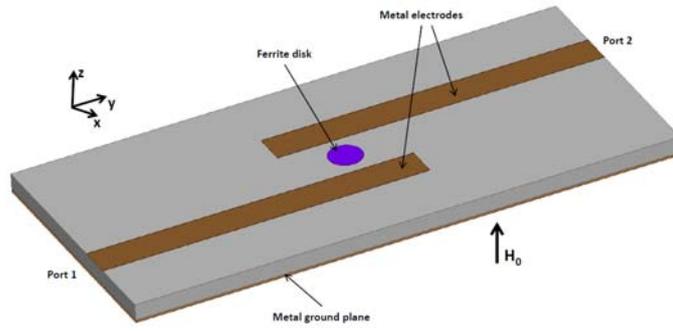

Fig. S4. A microstrip structure with an embedded quasi-2D ferrite disk.

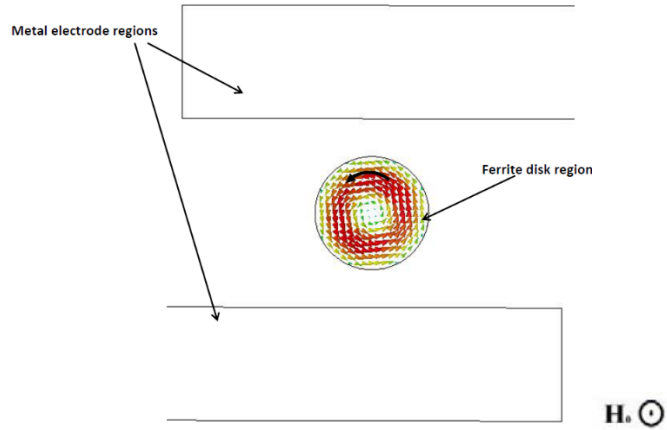

Fig. S5. Power-flow vortex for the 1st radial mode. Bottom plane of the disk. A black arrow clarifies the power-flow direction. A bias magnetic field is directed along $+z$ direction. The view is along $-z$ direction. The vortex is classified by the topological charge $Q=+1$.

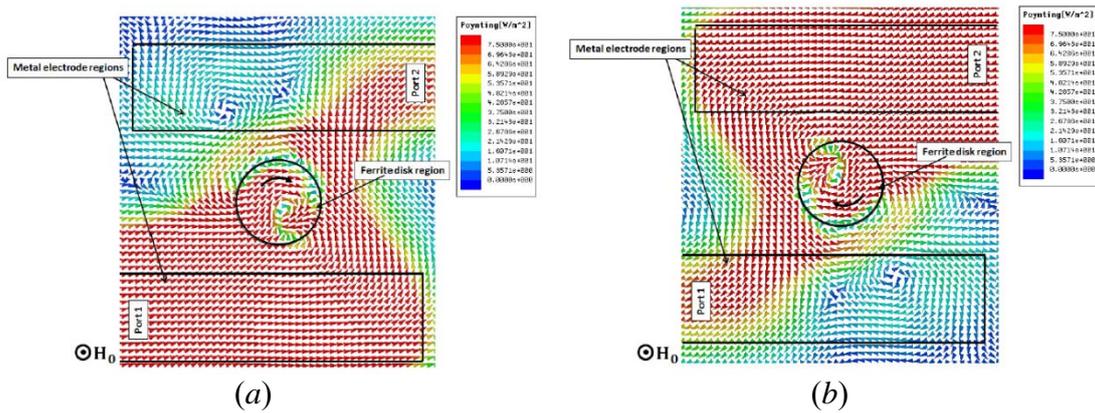

(*a*)                  (*b*)

Fig. S6. Power-flow vortex for the 1st radial mode on a ground plane of a microstrip structure. A black arrow clarifies the power-flow direction. A bias magnetic field is directed along $+z$ direction. The view is along $-z$ direction. The vortex is classified by the topological charge $Q=-1$. (a) Electromagnetic wave propagation in a waveguide from port 1 to port 2 (1 ➔ 2), (b) wave propagation from port 2 to port 1 (2 ➔ 1).

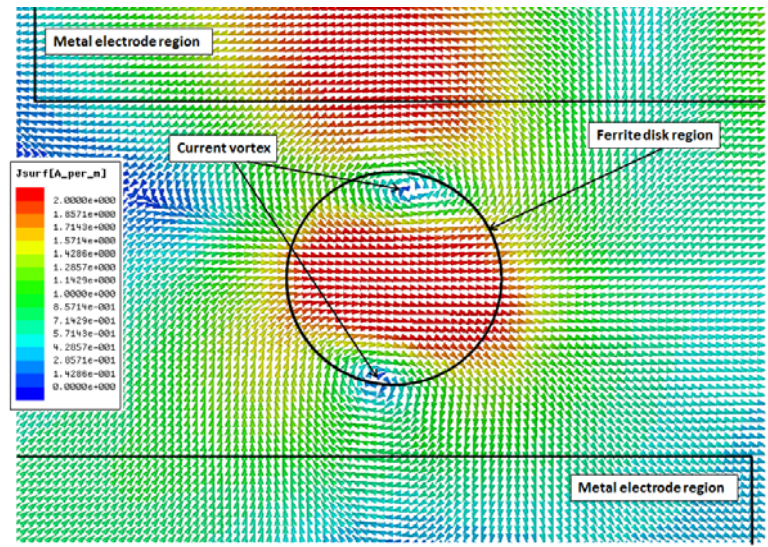

Fig. S7: Surface electric current for the 1st radial mode on a ground plane of a microstrip structure for a certain time phase.